\documentclass{aa} 

\newcommand{\RWBL}{WISE\,J141046.00$+$740511.2}
\usepackage{graphicx}
\usepackage{xcolor}
\usepackage[normalem]{ulem}

\usepackage{txfonts}
\usepackage{caption}
\usepackage{lineno}
\graphicspath{{figures/}}
\begin{document}

   \title{Disentangling the nature of the prototype radio weak BL Lac:}

   \subtitle{Contemporaneous multifrequency observations of WISE J141046.00$+$740511.2}

   \author{E. J. Marchesini\inst{1} 
            \and
            V. Reynaldi \inst{2,3}
            \and
            F. Vieyro\inst{3,4} 
            \and
             J. Saponara\inst{4} 
            \and
            I. Andruchow \inst{3,4}
            \and
            I. E. L\'opez \inst{1,5}
            \and
          P. Benaglia\inst{4}
            \and
          S. A. Cellone \inst{3,6}
            \and
          N. Masetti \inst{1,7}
            \and
          F. Massaro\inst{8,9}
            \and
          H. A. Pe\~na-Herazo\inst{10}
            \and
          V. Chavushyan \inst{11,12}
            \and
          J. A. Combi\inst{3,4,13}
          \and
          J. A. Acosta-Pulido\inst{14,15}
          \and
          B. Ag\'is Gonz\'alez\inst{16}
          \and
          N. Castro-Segura\inst{17}
             }

\institute{INAF -- Osservatorio di Astrofisica e Scienza dello Spazio, via Gobetti 93/3, I-40129, Bologna, Italy.
\and Instituto de Astrof\'isica de La Plata, CONICET--UNLP, CCT La Plata, Paseo del Bosque, B1900FWA, La Plata, Argentina.
\and Facultad de Ciencias Astron\'omicas y Geof\'isicas, Universidad Nacional de La Plata, Paseo del Bosque, B1900FWA, La Plata, Argentina.
\and Instituto Argentino de Radioastronom\'ia, CONICET-CICPBA-UNLP, CC5 (1894) Villa Elisa, Prov. de Buenos Aires, Argentina.
\and Dipartimento di Fisica e Astronomia "Augusto Righi", Università di Bologna, via Gobetti 93/2, 40129 Bologna, Italy.
\and Complejo Astron\'omico ``El Leoncito'' (CASLEO), CONICET-UNLP-UNC-UNSJ, San Juan, Argentina.
\and Instituto de Astrof\'isica, Facultad de Ciencias Exactas, Universidad Andr\'es Bello, Fern\'andez Concha 700, Las Condes, Santiago RM, Chile.
\and INFN -- Istituto Nazionale di Fisica Nucleare, Sezione di Torino, via Pietro Giuria 1, I-10125 Turin, Italy.
\and Dipartimento di Fisica, Universit\`a degli Studi di Torino, via Pietro Giuria 1, I-10125 Turin, Italy.
\and East Asian Observatory, 660 North A'oh{\=o}k{\=u} Place, Hilo, Hawaii 96720, USA.
\and Instituto Nacional de Astrof\'isica, \'Optica y Electr\'onica, Luis Enrique Erro \#1, Tonantzintla, Puebla 72840, M\'exico.
\and Center for Astrophysics | Harvard \& Smithsonian, 60 Garden Street, Cambridge, MA 02138, USA.
\and Departamento de F\'isica (EPS), Universidad de Ja\'en, Campus Las Lagunillas s/n, A3, 23071 Ja\'en, Spain.
\and Instituto de Astrof\'isica de Canarias (IAC), E-38205 La Laguna, Tenerife, Spain.
\and Departamento de Astrof\'isica, Universidad de La Laguna (ULL), E-38206 La Laguna, Tenerife, Spain.
\and Instituto de Astrof\'isica de Andaluc\'ia - CSIC, Glorieta de la Astronom\'ia, S/N, 18008 Granada, Spain.
\and Department of Physics \& Astronomy. University of Southampton, Southampton SO17 1BJ, UK.
}

 
  \abstract
   {The $\gamma$-ray emitting source \RWBL\ has been associated with a Fermi-LAT detection by crossmatching with Swift/XRT data. It has shown all the canonical observational characteristics of a BL Lac source, including a power-law, featureless optical spectrum. However, it was only recently detected at radio frequencies and its radio flux is significantly low.}
   {Given that a radio detection is fundamental to associate lower-energy counterparts to Fermi-LAT sources, we aim to unambiguously classify this source by performing a multiwavelength analysis based on contemporaneous data.}
   {By using multifrequency observations at the Jansky Very Large Array, Giant Metrewave Radio Telescope, Gran Telescopio Canarias, Gemini, William Herschel Telescope and Liverpool observatories, together with Fermi-LAT and Swift data, we carried out two kinds of analyses. On one hand, we studied several known parameters that account for the radio loudness or weakness characterization and their application to blazars (in general) and to our source (in particular). And, on the other hand, we built and analyzed the observed spectral energy distribution (SED) of this source to try to explain its peculiar characteristics.}
   {The multiwavelength analysis indicates that \RWBL\ is a blazar of the high-frequency peaked (HBL) type that emits highly polarized light and that is likely located at a low redshift. In addition, the one-zone model parameters that best fit its SED are those of an extreme HBL (EHBL); this blazar type has been extensively predicted in theory to be lacking in the radio emission that is otherwise typical of canonical $\gamma$-ray blazars.}
   {We confirm that \RWBL\, is indeed a highly polarized BL Lac of the HBL type. Further studies will be conducted to explain the atypical low radio flux detected for this source.}


   \maketitle
%

\section{Introduction}

Blazars are considered a sub-type of active galactic nuclei (AGN), whose relativistic jets are closely aligned to the line of sight \citep{BlandfordRees78,Lister13}. Their spectra are dominated by non-thermal emission over the whole electromagnetic range, namely, they can be detected at all radio frequencies \citep[see, e.g.,][]{Brown89,Giroletti16,Lister19}, infrared \citep[e.g.,][]{Impey88,Massaro11a,DAbrusco19}, optical \citep[e.g.,][]{Carini92,Marchesini16,alvarezCrespo16,Paiano20}, X-ray bands \citep[e.g.,][]{Singh85,Giommi90,Sambruna96,Paggi13,Marchesini19A}, and $\gamma$-ray band \citep[e.g.,][]{Hartman99,1LAC,Ackermann15,4LAC}. According to their optical spectra, \cite{Stickel91} classified blazars into two main categories. Broad optical emission lines, when present in their spectra, indicate they are categorized as flat-spectrum radio quasars (FSRQs), while sources that present weak or absent emission lines are considered as BL Lac objects \citep[see also][]{Urry95}. The spectral energy distributions (SEDs) of blazars exhibit two broad bumps, one at low energies and the other at higher ones. The low-energy bump is located between the infrared and X-rays bands and it is attributed to synchrotron emission arising from accelerated electrons in the blazar jet \citep[][]{Maraschi92, Tramacere07,Potter12}. The high-energy bump can be found between the hard X-rays and the $\gamma$-ray band, and its nature is still under debate: in leptonic models, it is attributed to the inverse Compton (IC) scattering of synchrotron photons (synchrotron-self Compton, SSC) or external photon fields (EC), such as those emitted in the Broad-Line region or in the accretion disk \citep[e.g.,][]{Ghisellini85,Tramacere07,Paggi09a}. In lepto-hadronic models, synchrotron radiation by protons and hadronic processes can also contribute to the high-energy component observed in blazars \citep{aha2000,mucke2001,atoyan2003,bottcher2013}. 

In $\gamma$-rays, in particular, blazars are the dominant species of detected sources \citep{Aharonian05,Abdo10A,Arsioli20}, having been also detected at extreme $\gamma$-ray energies \citep[i.e., TeV energies, see for example][]{WakelyHoran08}. They are also known for undergoing high-energy flaring states \citep[e.g.,][]{Hartman01,Kaur17,Bruni18}. In the Fourth Catalog of the the Fermi Large Area Telescope (4FLG, Fermi-LAT, \citealt{4FGL}), the collaboration lists 3130 blazars out of a total of 5064 sources ($\sim$62\%) detected above a 4$\sigma$ significance level. It is expected that a number of the 1336 Fermi-LAT objects still without any known lower-energy counterpart also belong to the blazar class \citep{Acero15}. Blazars are also known for being X-ray emitters as well \citep{Falcone14,Chang17}.

Different methods are currently available to look for blazar-like counterparts of $\gamma$-ray unidentified sources, which can later be confirmed as the true counterparts through their optical 
spectra \citep[see, e.g.,][]{Masetti13a,Marchesini16,PenaHerazo17,Marchesini19}, For example, $\gamma$-ray blazars populate a specific region in the color-color diagram at mid-infrared wavelengths \citep[see e.g.,][]{Massaro12,DAbrusco13}. In addition, Fermi-LAT blazars are classified with a multiwavelength approach. In particular, radio detection has been successfully used to identify blazars: in BZCAT \citep[][]{BZCAT2015}, all sources that were identified as blazars were also detected in radio frequencies, regardless of their high-energy detection status. This includes all blazars detected by Fermi-LAT, which has been the basis of the $\gamma$-ray to radio connection in these sources \citep{Mahony10,Ackermann11,Lico17}.

It was recently established that Fermi BL Lacs are also X-ray emitters \citep{Marchesini19A}. This has led to the selection of a sample of X-ray emitting blazar-like counterparts of Fermi unidentified sources displaying multiwavelength features similar to canonical blazars \citep{Marchesini20}. Among this list of blazar-like counterpart candidates for Fermi sources, several of them do not show any known radio detection to date. Such is the case of the Fermi BL Lac object \RWBL. This source was first suggested as a putative counterpart to a Fermi detection by \citet{Landi15}, which was then first confirmed as the true counterpart with a BL Lac optical spectrum by \citet{Marchesini16}. The fact that, at the time, there were no positionally-coincident radio detections led to its classification as the first radio weak BL Lac \citep[RWBL,][]{Massaro17}. Recently, \cite{Bruni18} proposed new RWBL candidates, although the subject has been under debate \citep{CAO}.

The existence of these sources was suggested by \citet{Ghisellini98}. These authors discussed the possibility of having BL Lacs with such a low intrinsic power that, according to the blazar sequence \citep{Fossati1998}, their synchrotron component should peak at frequencies above $10^{17}$ Hz, and the $\gamma$-ray component at TeV energies, thus burying the radio-band emission below present detection levels. Thus, they cannot be detected by large radio surveys, such as NVSS or SUMSS \citep[][]{Condon98,Mauch03}. This implies that the best way to find these sources is through their X-ray and $\gamma$-ray emission. There are several blazars detected with their synchrotron emission peaking at 2-10 keV, or even as high as 100 keV during flaring states \citep{Costamante01,Bonnoli15}. Regarding the classification in low-frequency peaked BL Lacs (LBLs) or high-frequency peaked BL Lacs (HBLs), defined by \citet{Padovani95}, these sources are known as Extreme HBL (EHBLs)\footnote{The term EHBL is also used to refer to BL Lacs extreme in $\gamma$-rays, which show intrinsic hard spectrum up to TeV energies \citep{Tavecchio11}.}. To date, however, none of the known EHBLs were radio-weak and detectable by Fermi-LAT at the same time \citep[see, e.g.,][and references therein]{Bonnoli15,MAGIC19}.

Thanks to a plethora of multifrequency contemporaneous observations, we aim to investigate the observational properties of \RWBL~ to confirm its nature, which could allow it  to serve as  the prototype of its class. The paper is organized as follows. In Sect. \ref{data}, we describe the observations and data reduction processes. In Sect.~3, we present our multifrequency analysis of \RWBL. In Sect. 4, we discuss various models for reproducing the data. We state our results and conclusions in Sect. \ref{conc}.

\section{Observations, data reduction, and results}\label{data}

We performed a multiwavelength observational follow-up campaign on WISE\,J141046.00+740511.2 \footnote{$\rm{RA}=14\rm{h}10\rm{m}46\rm{s}, \rm{DEC}=74\rm{d}5\rm{m}11.2\rm{s}$ (J2000.0).} 
between 2018-2019. A summary of the campaign is shown in Table~\ref{tab:Obs}, where we list the observatories used, the dates of each observation shift, the exposure time, and the observing band for each observation.

   \begin{table}
      \caption[]{Observing logs.}
         \label{tab:Obs}
     \centering
         \begin{tabular}{cccc}
            \hline
            \noalign{\smallskip}
            Observatory      &  Date & Exposure & Band \\
                             &       & Time [s] & \\
            \noalign{\smallskip}
            \hline
            \noalign{\smallskip}
            \hline
            Radio & & & \\
            \hline
            GMRT      & 21-Aug-2018  & 8520 & L-band \\
                      & 23-Aug-2018  & 8400  & 610 MHz \\
            JVLA      & 23-Nov-2018  & 2400   & L-band \\
                      & 23-Nov-2018  & 660  & S-band \\
           \noalign{\smallskip}
            \hline
            \noalign{\smallskip}
            \hline
            Optical Spect. &  & &\\
            \hline
            GTC & 13-Mar-2019  & 3600 & 400-900 nm \\
            Gemini    & 03-Apr-2019  & 1200 & 450-1000 nm \\
            \noalign{\smallskip}
            \hline
            \noalign{\smallskip}
            \hline
            Optical Phot. & & & \\
            \hline
            Liverpool & 04-Feb-2019 & 50 & SDSS-g' \\
                      &             & 150 & SDSS-r'\\    
                      &             & 150 & SDSS-i' \\               
            WHT & 29-Apr-2019 & 30 & SDSS-G \\
                &             & 30 & SDSS-R \\
                &             & 25 & SDSS-I \\
           \noalign{\smallskip}
            \hline
            \noalign{\smallskip}
            \hline
            Ultraviolet & & & \\
            \hline
  Swift/UVOT &  2012-2014 & 11700 &  u\\         
 &   & & uvw1 \\         
  &   & &  uvm2\\         
  & &  &  uvw2\\         

            \noalign{\smallskip}
            \hline
            \noalign{\smallskip}
            \hline 
            X-rays & & & \\
            \hline
            Swift/XRT &  2012-2014 & 11700 & 0.5-10 keV \\
            \noalign{\smallskip}
            \hline
         \end{tabular}
   \end{table}

\subsection{Radio observations}

\subsubsection{Giant Metrewave Radio Telescope observations}

The Giant Metrewave Radio telescope (GMRT) observations were conducted during August 2018 in two different frequency bands centred at 1260~MHz and 607~MHz, considering the back-end non-polarimetric-configuration 400~MHz and 200~MHz bandwidths, respectively. The flux calibrator source, 3C~286, was observed at the beginning of both runs for $\sim$8~min\footnote{All the exposure times reported for every source observed at radio frequencies are dwell times.} at each frequency band, while the source 1407$+$284 was used as the phase calibrator and was observed for $\sim$5~m between the target scans of $\sim$30~m on \RWBL\ (see Table \ref{tab:Obs} for details). The data were flagged and calibrated using ``A flagging and calibration pipeline for GMRT data'' \citep[FLGCAL,][]{2011ascl.soft12007P}, while the imaging process was carried out with the Common Astronomy Software Applications \citep[CASA,][]{CASA-2022}. The images were produced considering a robust weighting of 0.5, and discarding baselines shorter than 2.1~km and 4.9~km, at 1.4~Ghz and 610~Mhz respectively. In addition, for further analysis and visualization imaging, we extensively
used the {\sc miriad} software package \citep{1995ASPC...77..433S} and {\sc kvis}, part of the {\sc karma} package \citep{1996ASPC..101...80G}. Due to the vast presence of radio frequency interference (RFI) during the observations at the L-band (1.4 GHz), along with the presence of a strong radio source close to \RWBL, the required root mean square (rms)  to detect the source in this band was not achieved.  The synthesized beam and r.m.s. (at the center of the field) obtained at 610~MHz are $7\farcs8\times 3\farcs4$ and 0.1 mJy~beam$^{-1}$. In this band, the source is a point source with a flux density of $2.1\pm0.1\,\rm{mJy}$, which is in agreement with previous detections in the literature \citep[][]{Schinzel17}.%

\subsubsection{Karl G. Jansky Very Large Array observations}

\begin{figure}
    \centering
    \includegraphics[width=0.4\textwidth]{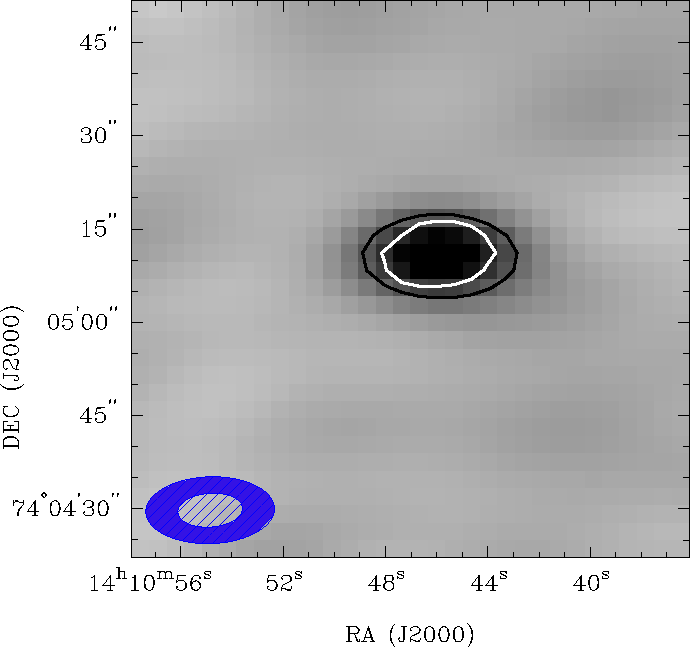}\\
\caption{JVLA L-band map of \RWBL. The L-band (1.4 GHz) continuum image is overlaid with L-band (black) and S band (3 GHz, white) contours. Contour levels are drawn at a 3$\sigma$ image noise level, where $\sigma$ is 0.3~mJy~beam$^{-1}$ and $\sigma$ is 0.1~mJy~beam$^{-1}$ in the L and S-bands, respectively. The synthesized beams of L and S-bands are shown at the bottom left corner of the image, $23\farcs85\times 12\farcs63$ and $10\farcs32\times 5\farcs3$, respectively.}
\label{fig:vla}
    \end{figure}

\RWBL\ was observed in November 2018 with the the Karl G. Jansky Very Large Array (JVLA) in C configuration. The observations were carried out at two different frequency bands, centred at 1420~MHz and 3000 MHz, using a configuration of 1~GHz and 2~GHz bandwidths, respectively. The flux calibrator source 3C286 was observed at the beginning of the run for $\sim$3~min at each frequency band. The source 1459$+$7140 was used as the phase calibrator and was observed for $\sim$2 min between the target scans; see Table~\ref{tab:Obs} for details.
The data were calibrated and the image was performed in the standard way using CASA. The images were built considering a robust weighting of 0.5. The synthesized beams are $23\farcs85\times 12\farcs63$ and $10\farcs32\times 5\farcs3$ with an rms attained at the field center of 0.3~mJy~beam$^{-1}$ and 0.1~mJy~beam$^{-1}$ at the frequencies of 1.4 and 3.0~GHz, respectively (see Fig.~\ref{fig:vla}). The source is a point source with a flux density of $2.33\pm0.65\,\rm{mJy}$ at 1420~MHz, and $2.36\pm0.60\,\rm{mJy}$ at 3000 MHz.

\begin{table}
    \caption{VLA \& GMRT data.}\label{tab:fluxes}
\begin{tabular}{cccccc}
\hline
 $\nu_0$ & $\Delta \nu$ & $F_{\rm VLA}$ & $F_{\rm GMRT}$& $\alpha_{\rm GMRT}$ & $\Delta \alpha$ \\
 
 [GHz] & [GHz] & [mJy] & [mJy] & & \\
\hline
  0.610 & 0.033 &              & 2.1$\pm$0.1 &      &    \\
  1.42  & 0.064 &  2.33$\pm$0.65 &   --          &   &  \\
  3.0   & 0.128 &  2.36$\pm$0.60 &             &-0.11 & 0.07 \\  
\hline
\end{tabular}\\
\end{table}

\subsection{Optical observations}

\subsubsection{GTC observations}\label{gtc}

We acquired a long-slit optical spectrum with the OSIRIS spectrograph at Gran Telescopio de Canarias (GTC). Observations were carried-out using queue mode on March 13, 2019 at 05:30 UT time. We used the R300B grism in the spectral range of 400-900 nm, with a dispersion of 0.496 nm/pix. We acquired three observations, with a exposure time of 1200 seconds each.

\begin{figure}
\centering
\includegraphics[width=0.55\textwidth]{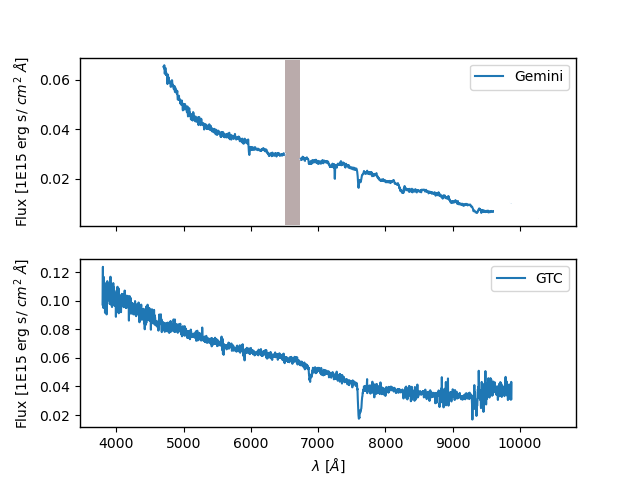}
\caption{Optical spectra of \RWBL. Top panel: Spectrum taken with Gemini GMOS-N. Bottom: Spectrum taken with GTC-OSIRIS. In both cases, all absorption features are either due to atmospheric absorption (telluric lines), or instrumental artifacts (such as the CCD gap in the GMOS-N data).}
\label{fig:specgemini+gtc}
\end{figure}

We reduced the spectroscopic data using standard procedures with the IRAF\footnote{\textit{Image Reduction and Analysis Facility}} package \citep{Tody86}. We performed bias subtraction and flat-field correction using dome flats. We also removed cosmic rays using L.A. Cosmic IRAF algorithm \citep{vanDokkum01}. We used HgAr Arc lamps for the wavelength calibration. We show the reduced spectrum in Fig.~\ref{fig:specgemini+gtc}. No spectral features are apparent in the resulting spectrum.

\subsubsection{Gemini-N observations}\label{gemini}

The object \RWBL\ was observed with the GMOS \citep{Hook04} instrument of Gemini North Telescope on April 4th 2019 (GN-2019A-Q-116; PI: E. Marchesini). The instrument  was set up in longslit mode, with a slit of 1 arcsec width. We chose the R150-G5308 grating, which yields a dispersion of 0.174 nm per pixel. This grating was used with the aim of obtaining the largest available spectral coverage in the optical band. The data were obtained in two separate exposures of 600 s each, centered on two different wavelengths (520 nm and 545 nm) to correct for the gaps in the GMOS detector.

The reduction procedure included the usual steps of bias subtraction, flat-field correction, wavelength calibration, sky subtraction, and cosmic ray rejection by using the {\sc gemini.gmos} package reduction tasks (v1.14) within IRAF (v2.16). The resulting spectrum covers from 450 nm to 1000 nm. No emission lines were detected, reinforcing the classification of this source as a blazar.

\subsubsection{Liverpool Telescope observations}\label{liv}

\RWBL\ was observed with the IO:O camera equipped with a e2V CCD 231-84, at the Liverpool Telescope\footnote{\url{https://telescope.livjm.ac.uk/TelInst/Inst/IOO/}}, Roque de los Muchachos Observatory, in La Palma, Canary Islands. The observation was made using the $g'$, $r'$ and $i'$ Sloan filters system, with an exposure time of 50 seconds in the $g'$ filter, and 150 seconds in the  $r'$ and $i'$ filters; on February 4, 2019 (corresponding to the night of February 3). The image scale was 0.3037 arcsec\,px$^{-1}$ with $2 \times 2$ binning, and the seeing during the night was between 1.57 and 2.01 arcsec.

\begin{figure}
\centering
\includegraphics[width=0.47\textwidth]{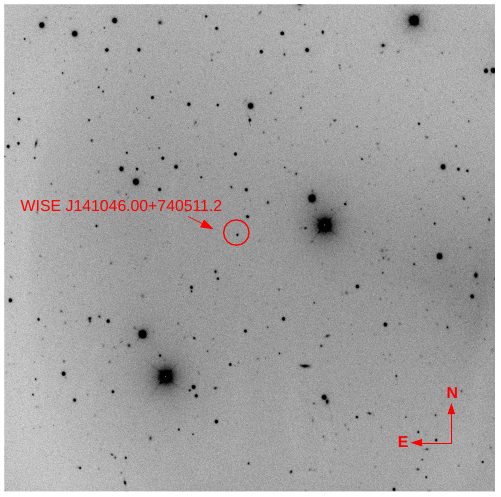}
\caption{Image of the field of WISE J141046.00+740511.2 taken with the IO:O camera of the Liverpool Telescope. The image corresponds to the g' filter of the SLOAN filter system and the image size is 10'x10'.}
\label{fig:liverpool}
\end{figure}

Standard procedures were applied to reduce the data, subtracting the bias and correcting for flat fields. Aperture photometry was performed using the \texttt{apphot} IRAF software routine. Since we obtained six images in $g'$, 5 in $r'$, and 11 in $i'$, we also performed differential photometry, searching for possible signs of variability at very short scales. We did not find any variability. We performed a flux calibration using a reference star in the same field from the Pan-STARRS1 catalog \citep[][]{Flewelling2020}. The mean standard magnitude values for \RWBL\ were: $m_{\rm{g'}}=19.977\pm0.053$, $m_{\rm{r'}}=19.627\pm0.022$, and $m_{\rm{i'}}=19.357\pm$0.026. Corresponding flux values are given in Table \ref{tab:data}. As an example, we show an image in the g' band built with these data in Fig.~\ref{fig:liverpool}

\subsubsection{William Herschel Telescope observations}

The target was also observed using the William Herschel Telescope (WHT), Roque de los Muchachos Observatory, in La Palma, Canary Islands. We used the Auxiliary-port CAMera (ACAM\footnote{\url{https://www.ing.iac.es//Astronomy/instruments/acam/}}) at image mode, in the $G$, $R$ and $I$ Sloan filter system on April 29, 2019. The exposure times were 30, 30, and 25 seconds in each filter, respectively. The data were reduced and the photometry performed using the same process described in Sect. \ref{liv}. The magnitudes resulted in very similar values to the ones obtained from the Liverpool data ($m_{\rm{g}}=19.832 \pm 0.017$, $m_{\rm{r}}=19.649 \pm 0.016$, and $m_{\rm{i}}=19.58 \pm 0.047$). The corresponding flux values are shown in Table \ref{tab:data}. We note that the filter systems used by both telescopes have small effective wavelength differences, which may explain the different (at the $\sim 0.1$\,mag level) magnitude values obtained.

We also used the Intermediate-dispersion Spectrograph and Imaging System (ISIS) in image mode to obtain linear polarization. The source was observed on 2019-04-29 with the $V$ 
Johnson filter, and with half-wave plate angles of 8$^{\circ}$, 30.5$^{\circ}$, 53$^{\circ}$, and 75.5$^{\circ}$. The zero-polarization standard star used was BD+332642 \citep{Turnshek90}. 

For the reduction, we subtracted the bias and applied a flat-field correction using sky flats and masking the gaps on the CCD, produced by the vignetting of the instrument. We carried out aperture photometry for our source and the standard star. After subtracting the sky level, we obtained the fluxes for both filters and the four wave-plate angles positions. Using the equations given in \citet{Zapatero05}, we derived the linear optical polarization degree and its uncertainty in the V band. After the zero-polarization correction, we obtained a value of P$_V$=7.07$\pm$2.12\%. The level of uncertainty is due to the weakness of the source. Nonetheless, we note that \RWBL\, shows a polarization that is several degrees ($>5$) higher than all the rest of the sources in the same field.

\subsection{Swift observations}

\subsubsection{Swift UltraViolet and Optical Telescope (UVOT)}

\RWBL\ was observed by Swift for a total of 11.7 kiloseconds (hereafter,  "ks"), on different dates from 2012 to 2014. In particular, we found a total of 20 UVOT exposures on the source, in filters u, w1, m2, and w2.

We followed the basic standard procedures to reduce Swift/UVOT data with HEASOFT tools, as described by the UVOT Analysis Threads from the University of Leicester\footnote{\url{https://www.swift.ac.uk/analysis/uvot/}} and in previous analyses \citep[see e.g.,][]{Tramacere07b,Massaro08b,Paggi13,Maselli16}. In the following, we provide a brief overview.

We first merged together all the observations of the same filter with the \texttt{uvotimsum} task, as well as all their corresponding exposure maps. We then used the \texttt{uvotdetect} task on the merged images, using their merged exposure maps, to find all sources in the field with signal-to-noise ratios (S/N) greater than 3. We performed accurate photometry on \RWBL\ with the \texttt{uvotsource} task, on a circular region with a 5 arcsec radius centered on the position returned by \texttt{uvodetect}. The background extraction region was defined as an annulus centered on the same position, with a much larger area which does not include the source region nor any other detected sources. The obtained flux is shown in Table \ref{tab:data}.

\subsubsection{Swift X-ray Telescope (XRT)}

We followed the same Swift/XRT data reduction basic procedure as described in previous works \citep[see e.g.,][]{Massaro11b,Massaro17,Marchesini19A,Marchesini20}. Clean event files were obtained using the {\sc xrtpipeline} task of the Swift X-ray Telescope Data Analysis Software package \citep[{\sc xrtdas},][]{Capalbi05}. All time intervals exceeding 40 counts per second were excluded, as well as those during which the CCD temperature exceeded -50$^{\circ}$\,C in edge locations on the CCD \citep{DElia13}. The total exposure time was of 11.7 ks. The source was detected with a countrate of $ (2.8 \, \pm \, 0.6)\, \rm{E-3} \, \rm{cts/s}$, in the 0.5-10 keV band. The corresponding flux was obtained using {\sc pimms} \citep{Mukai93}, assuming a power-law function with a Photon index of 2, and a galactic nH of 0.0225. The resulting flux is shown in Table \ref{tab:data}.

\subsubsection{Catalog data: WISE and Fermi-LAT}

We chose to add the mid-IR and $\gamma$-ray data available online\footnote{\url{https://wise2.ipac.caltech.edu/docs/release/allwise/}}\footnote{\url{https://fermi.gsfc.nasa.gov/ssc/data/access/lat/10yr$\_$catalog/}} for completeness, since these bands were crucial in the classification of this source \citep[see][]{Massaro17}. The mid-IR data, from WISE \citep{Wright10}, correspond to the 3.4, 4.6, 12, and 22 $\mu$m bands, taken in 2010 (see Sect. 4.). On the other hand, the $\gamma$-ray data were taken from the following Fermi-LAT catalogs: Fermi Large Area Telescope First Source Catalog \citep[1FGL,][]{1FGL}, First Fermi-LAT Catalog of Sources above 10 GeV \citep[1FHL,][]{1FHL},  LAT 4-year Source Catalog \citep[3FGL,][]{3FGL}, Third Fermi-LAT Catalog of High-Energy Sources \citep[3FHL,][]{3FHL}, and the aforementioned 4FGL \citep{4FGL}. These data correspond to the energy bands of 100 MeV – 100 GeV (1FGL, 3FGL, and 4FGL), and of 10 GeV – 30 GeV (1FHL, 3FHL).

   \begin{table}[ht]
      \caption[]{Multiwavelength data used in this paper. Uncertainties are reported at the 1$\sigma$ level.}
         \label{tab:data}
     \centering
     \fontsize{9}{11}\selectfont
         \begin{tabular}{rr}
            \hline
             \hline
            Frequency       & Flux $\pm$ Uncertainty   \\
            $[\textrm{Hz}]$   & [erg cm$^{-2}$ s$^{-1}$]\\
            \hline
            \hline
            GMRT/VLA & \\
            \hline

            6.1E8       &       1.281 $\pm$ 0.003   \,E-18\\
            1.4E9       &       3.31  $\pm$ 0.04    \,E-18\\
            3.0E9       &       7.08  $\pm$ 0.08        \,E-18\\

            \hline
            \hline
            WISE &\\
            \hline

            8.8\,E13    &       2.7      $\pm$ 0.2 \,E-13\\
            6.5\,E13    &       2.3  $\pm$ 0.3 \,E-13\\
            2.5\,E13    &       2.7  $\pm$ 0.9 \,E-13\\
            1.3\,E13    &   10.2 $\pm$ 3.8 \,E-13\\

            \hline
            \hline
            WHT  & \\
            \hline

            6.3\,E14    &2.67 $\pm$ 0.04         \,E-13\\
            4.8\,E14    &2.41 $\pm$ 0.04         \,E-13\\
            3.9\,E14    &2.08 $\pm$ 0.09         \,E-13\\

            \hline
            \hline
            LIV  & \\
            \hline

            6.3\,E14    &       2.3  $\pm$       0.1 \,E-13\\
            4.8\,E14    &       2.45  $\pm$      0.05 \,E-13\\
            3.9\,E14    &       2.55  $\pm$      0.06 \,E-13\\

            \hline
            \hline
            Swift/UVOT  & \\
            \hline

            8.6\,E14    &       1.33  $\pm$  0.09 \,E-13\\
            1.1\,E15    &       1.2  $\pm$      0.1 \,E-13\\
            1.3\,E15    &       0.9  $\pm$ 0.1 \,E-13\\
            1.5\,E15    &       1.3  $\pm$  0.1 \,E-13\\

            \hline
            \hline
            Swift/XRT  & \\
            \hline

            1.2\,E18    &1.1 $\pm$ 0.2 \,E-13\\

            \hline
            \hline
            Fermi-LAT  & \\
            \hline
    
            4.8\,E24    &2.6 $\pm$      1.3 \,E-12\\
                &2.8 $\pm$      1.1 \,E-12\\
            1.2\,E25    &6.6 $\pm$      3.1 \,E-12\\
                &7.1 $\pm$      0.7 \,E-12\\
                &3.6 $\pm$      0.5 \,E-12\\
    
            \noalign{\smallskip}
          
            \noalign{\smallskip}
            \hline
         \end{tabular}
     
   \end{table}

\section{Multifrequency properties of \RWBL}

\subsection{On the optical classification}

Blazars of the BL Lac type are among the most elusive high-energy emitting objects to pinpoint \citep[][]{Massaro15c,Massaro16b,PenaHerazo20}. Their collimated jets, whose material is accelerated to relativistic velocities, point towards the line of sight. Thus, they suffer the effect of relativistic Doppler boosting, which dramatically increases their observed flux. This, in turn, overshines most of the stellar continuum (plus absorption lines) from the host galaxy, as well as any AGN emissions. The non-detection of optical spectral features is one of the defining criteria to classify any object as a BL Lac \citep[][]{Stickel91}.
The non-thermal origin of the emission in blazars generates, in the optical band, a blue, power law-like spectrum, which in itself is another typical signature of a BL Lac nature \citep[][]{Marcha96}. Non-periodical variability on very short time-scales, polarized emission, and association to high energy and/or very high energy sources are further criteria that are useful for classifying an object as a BL Lac, although this sort of evidence is generally considered less direct \citep{Laurent98}. 

To further confirm or discard the conclusion, as stated in \citet{Massaro17}, that the source \RWBL\ is indeed a blazar of the BL Lac type, we aim to present new, detailed optical spectra. In their work, \citet{Marchesini16} found all the typical characteristics of a BL Lac object in a spectrum of \RWBL\ taken with the Galileo National Telescope (TNG). They were not able to find any spectral features, although they indicate a hint of a possible emission line that is barely within detection.

However, both our spectra from Gemini-N and from GTC failed to detect any spectral feature whatsoever (see Fig. 2). More specifically, both these spectra confirm the fact that the possible emission hint found by \citet{Marchesini16} was, in all probability, an artifact, and not an actual feature from the source itself. We rule out variability on the spectrum since the source magnitude was constant during the whole period covered by all three observations. It is also worth noting that the sensitivity and resolution of both these telescopes in the configurations we used (see Secs. \ref{gtc} and \ref{gemini}) exceed the performance capabilities of TNG.

Moreover, taken together, our GTC and Gemini-N spectra cover the whole optical band, from $\sim 400\rm{nm}\,$ to $\sim 1000 \rm{nm}\,$. We did not detect any emission lines in this range. This could hint at the source lying at a relatively moderate distance, since it is simpler to swamp up intrinsically less intense emission lines \citep[see, e.g.,][]{Blandford90}.

Perhaps the most compelling, definitive evidence of the BL Lac nature of \RWBL\, is the fact that it also shows a high polarization degree (see Sect. 2.2.4). Such polarization degree is only possible for non-thermal emission processes, which, in turn, are dominant in only a handful of sources \citep[see][]{Massaro17}. However, the blue, power-law featureless spectrum paired with high polarization serves as the benchmark for the definitive classification of a source as a BL Lac \citep[see, e.g.,][]{Urry95,Laurent98}.

\subsection{Radio loudness}\label{radioloudness} 

Different parameters are used to determine the radio loudness of a given source. We discuss about these parameters and the way in which they are defined in the following, followed by a description of our probe into their relationship with our source and its nature.

\subsubsection{A radio criterion}

\RWBL\ is clearly detected in our JVLA observations at 1420~MHz and 3000~MHz, and in our GMRT pointings at 610~MHz (see Fig.~\ref{fig:vla}). We derived fluxes from these observations by fitting a Gaussian profile with the width of the synthesized beam, for each mentioned frequency (see Table~\ref{tab:fluxes}).

This source was first detected in radio wavelengths by \citet{Schinzel17}, and later by \citet{CAO}; both detections were published after the first claim of \RWBL\ as a RWBL source \citep{Massaro17}. In all cases, at 1.4 GHz, the source was reported to be detected with a flux of $\sim$2 mJy, which is always compatible within the uncertainties related to the detection reported in this work.

After \citet{gregg-1996}, the monochromatic radio power at $1.4$~GHz (i.e. $L_{\rm 1.4~GHz}$) sets the threshold to separate radio-loud from radio-weak quasars, in such a way that radio-loud sources are characterized by $L_{\rm 1.4~GHz}$>10$^{25.5}$~W~Hz$^{-1}$ (or $L_{\rm 1.4~GHz}>4.4\times 10^{41}$~erg~s$^{-1}$). The monochromatic radio power is defined as:

\begin{equation}\label{eq:L}
L_{\rm 1.4~GHz} = \frac{S_{\rm 1.4} 4\pi D^2}{(1+z)^{1+\alpha}},
\end{equation}

\noindent where $D$ is the luminosity distance\footnote{
We adopt a flat cosmology with $H \sim 70$~km~$\rm s^{-1} Mpc^{-1}$, $\Omega_{\Lambda}$=0.7, $\Omega_{m}$=0.3; and 1~mJy=$10^{-29}$~[W\,$\rm m^{-2}\,Hz^{-1}$].}, $S_{1.4}$~[W\,$\rm m^{-2}\,Hz^{-1}$] is the observed flux, $z$ is the redshift, and $\alpha$ is the spectral index.

\begin{figure*}
    \centering

    \includegraphics[width=0.75\textwidth]{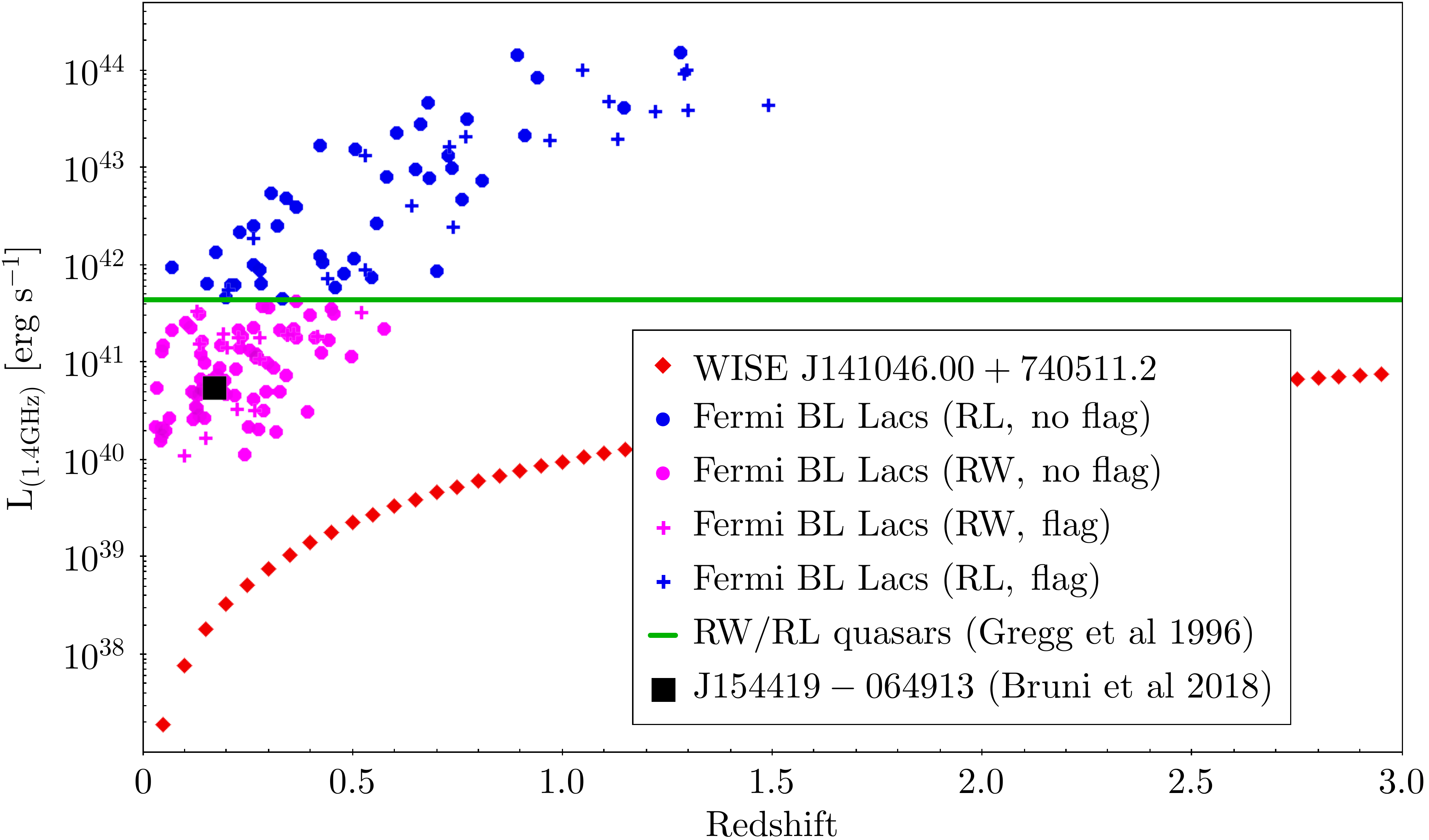}
    \includegraphics[width=0.75\textwidth]{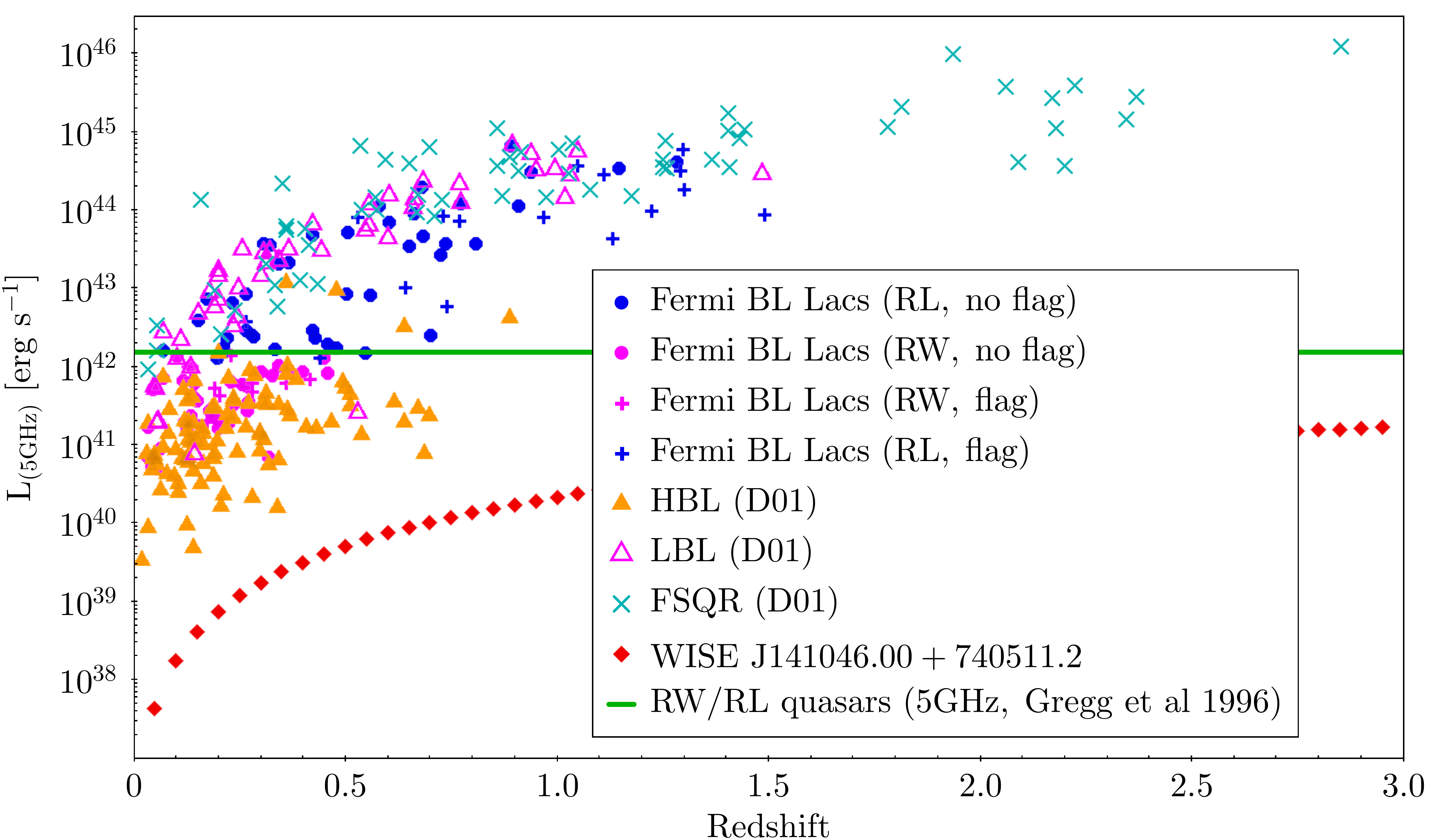}
\caption{ Luminosity of \RWBL\ at 1.4GHz as a function of redshift (red dotted line). The threshold of \citet{gregg-1996} (green solid line) separates the BZCAT sample of known-redshift BL~Lac sources into radio-loud (blue) and radio-weak (magenta) blazars. The red dotted line (diamonds) shows the luminosity that \RWBL\, would have as a function of redshift according to our 1.4 GHz data. The black square shows the position of the confirmed RWBL J154419$-$064913. In the lower panel, we show a comparison of the extrapolated luminosity at 5~GHz of our source (red dotted line) with the sample of HBL (orange filled triangles), LBL (open pink triangles), and FSRQ (cyan crosses) blazars.}
\label{fig:Lumin}
    \end{figure*}

In order to use $L_{\rm 1.4~GHz}$ to analyze our source, we selected all the available Fermi-detected BL~Lac objects with known redshift, from BZCAT \citep{BZCAT2015}. We included all objects with a redshift value, differentiating those with or without redshift detection flags, for a total of 146 blazars. The upper panel of Fig.~\ref{fig:Lumin} shows the $L_{\rm 1.4~GHz}=4.4\times 10^{41}$~erg~s$^{-1}$ threshold with a green, solid, horizontal line. Above this line, we find sources that should be classified as radio-loud BL~Lac sources (blue filled circles and crosses); and below the line, the radio-weak BL~Lac sources (magenta filled circles and crosses). Hereafter, we are going to refer to these subsamples as the radio-loud or radio-weak BL\,Lac samples, respectively, according to this criterion. The red dotted (diamonds) line represents the $L_{\rm 1.4~GHz}$ that our source would be expected to have as a function of $z$, according to Eq.~\ref{eq:L} and our 1.4~GHz VLA data (Table~\ref{tab:fluxes}). These points were obtained using the online tool provided by \citet{Wright06}. 
In order to compare both the method and our source with another confirmed RWBL, we used the data of \citet{Bruni18} to plot the RWBL source named J154419-164913 as a (black) filled square. This source is located where $L_{\rm 1.4~GHz}$ predicts to find the radio-weak sources, according to \citet[][]{gregg-1996}, and in spite of its non-quasar nature. The plot shows that \RWBL\ is a radio-weak source, independently of its redshift value.

To add another perspective, we have also included the classified sample of 209 blazars with known redshifts from \citet[][hereafter D01]{Donato01} in the lower panel of Fig.~\ref{fig:Lumin}. This sample allowed us to compare the $L$ threshold criterion with the spectral classification of the sources, since these sources have been classified into HBLs, LBLs, and FSRQs. Since the original data from D01 comprises the monochromatic power at 5~GHz, we changed the ordinates in the plot to take into account the luminosities of both samples at this frequency. The $L_{\rm 1.4~GHz}$ threshold defined by \citet{gregg-1996} was empirically translated to this new plot: we derived the tentative $L_{\rm 5~GHz}$ value ($1.5\times 10^{42}$ erg\, s$^{-1}$), which separates the sample in the same subsamples as the threshold in 1.4 GHz. The plot also includes the BZCAT sample of sources with both known redshift and detected radio flux at 5~GHz. We have kept their colours separating radio-weak (magenta) and radio-loud (blue) sources in order to emphasize that these subsamples were defined on the basis of their luminosity at 1.4~GHz. We note that the BZCAT criteria to define a source as a BL Lac is stricter than the criteria used by D01, which means that the latter may suffer from contamination.

Following the same procedure as in the upper panel, the red curve shows the luminosity of our BL~Lac source as a function of redshift.
The flux at 5~GHz was extrapolated from the 3~GHz data by using the spectral index of $\alpha=-0.11\pm0.07$ (Table~\ref{tab:fluxes}).

It is straightforward to conclude that the target is a radio-weak source. Irrespective of redshift, \RWBL\, shows a radio luminosity well below most known Fermi-LAT BL Lacs. In addition, \RWBL~ shares characteristics with low-redshift HBL sources. Moreover, Fermi-LAT detections of BL Lacs with fluxes and photon indices such as those of \RWBL\, are severely restricted in redshift due to pair-production extinction \citep[see, e.g.,][and references therein]{Kneiske04,Desai17}, which is consistent with the low-redshift HBL classification of this source.

\subsubsection{Comparing radio and optical powers}

\begin{figure*}
    \centering
    \includegraphics[width=0.75\textwidth]{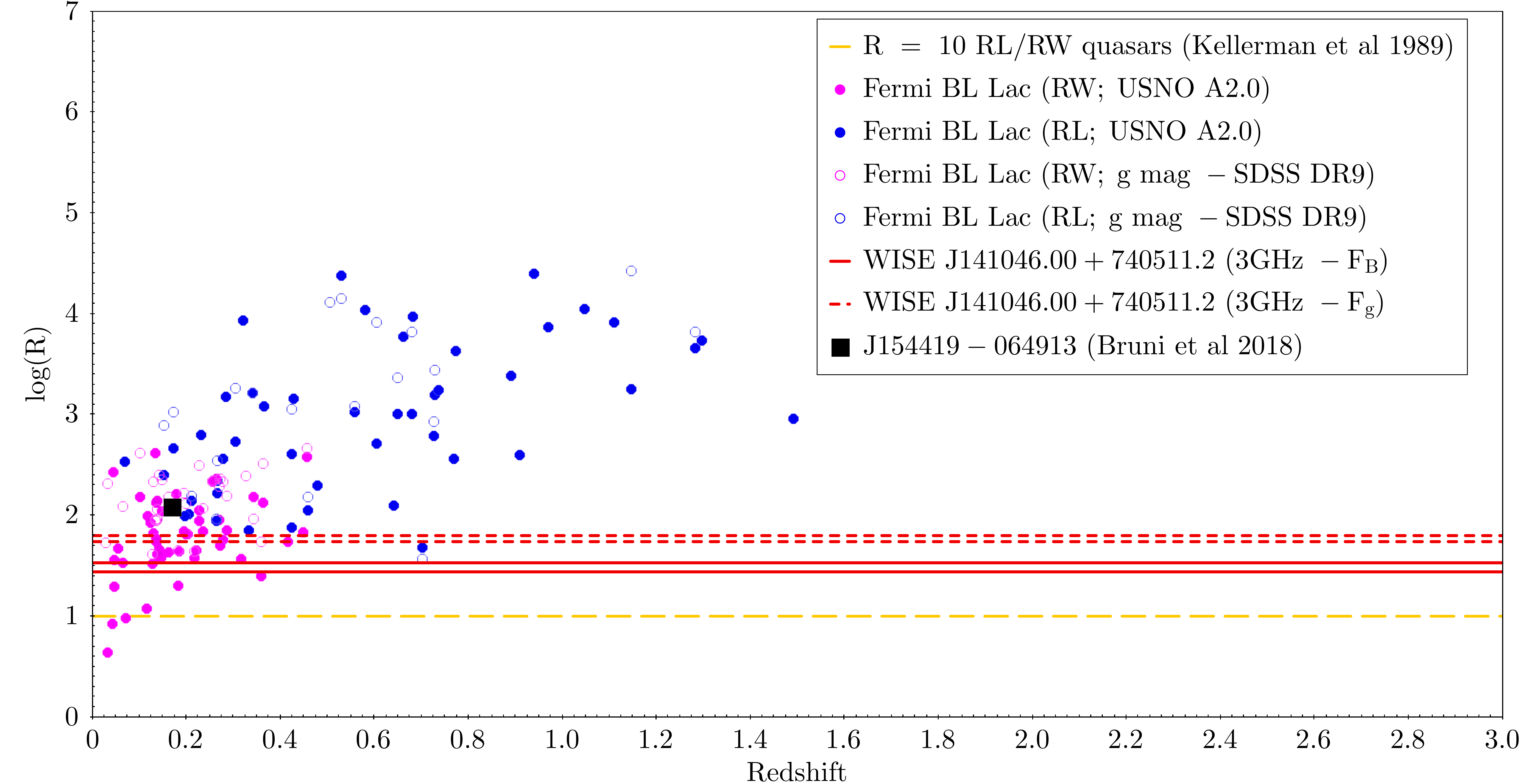}
    \includegraphics[width=0.75\textwidth]{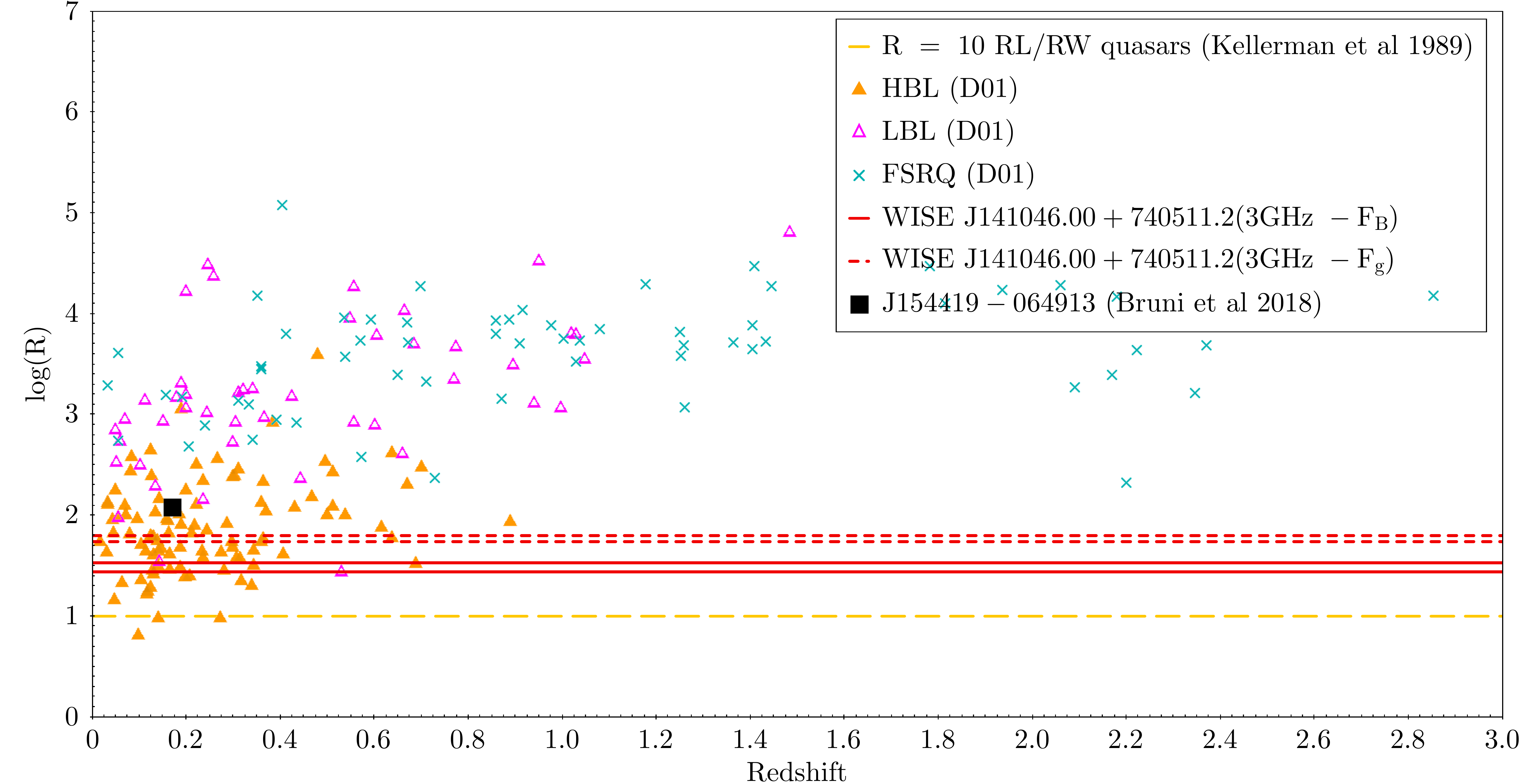}
\caption{$R$ parameter as a function of redshift. Data for \RWBL~ are shown in red. $R$ was calculated by using: B-fluxes from USNO and Gaia (red continuous lines) and $g'$-fluxes from WHT and Liverpool Telescope (the red dashed lines). {\it Upper panel}: BZCAT subsamples (set on the basis of their $L_{\rm 1.4~GHz}$; see Fig.~\ref{fig:Lumin}). Filled (empty) circles denote that $R$ was calculated by using B-fluxes from USNO~A2.0 ($g'$-fluxes from SDSS~DR9) data. {\it Lower panel}: Comparison of $R$ values between different blazar-types and \RWBL. }
\label{fig:R_z}
    \end{figure*}

Another parameter commonly used to determine the radio loudness of a given source is $R$, given by:

\begin{equation}\label{eq:R}
R = \frac{f_{5\textrm{ GHz}}}{f_{4400~\AA}},
\end{equation}

\noindent as proposed by \citet{GopalKrishna86}, \citet{Kellermann89}, and \citet{Schneider92} with the aim of classifying quasars. We used the same samples of BZCAT and D01 to analyze the reliability of $R$ in the context of blazars. Figure~\ref{fig:R_z} shows $R$ as a function of redshift. Both panels show the $R=10$ threshold ($\log(R)=1$; yellow, dashed line) that separates radio-loud ($\log(R)>1$) from radio-weak ($\log(R)<1$) quasars. The bottle-neck in the calculus of $R$ is the availability of optical data at 4400\AA (440 nm), both for our blazar samples and for \RWBL. The source was observed with WHT and Liverpool Telescope for this purpose, from where we obtained the fluxes in the $g'$ filter (SDSS filter system). These data are plotted as red short-dashed lines ($\log(R)_{WHT}=1.74$, $\log(R)_{Liv}=1.8$). We have also used the $B$ flux listed in the USNO\,A2.0 catalog \citep{usnoa2}, and the $B_p$ flux listed in the Gaia DR2 catalog \citep{gaiadr2}, which are plotted as red continuous lines ($\log(R)_{USNO}=1.44$, $\log(R)_{Gaia}=1.53$). We note that in both cases, our 3 GHz data were used instead of 5GHz in the calculus of $R$. For the sake of comparison, the BZCAT subsamples (as defined in the former paragraph, keeping their respective colours) are used with both optical filters in the upper panel of Fig.~\ref{fig:R_z}: filled symbols denote that $R$ was calculated by using the flux in the $B$-filter, as it is defined; whereas open symbols denote that the flux in the $g'$-filter was used instead of $B$ in Eq.~\ref{eq:R}. As it can be seen, differences in the usage of $B$ or $g'$ are indistinguishable. In the lower panel, we have plotted the D01 classified sample, and $R$ was calculated by using the 5GHz and 5500\AA~ fluxes from the same work. Both in the upper and lower panels we show the location of the known RWBL J154419$-$064913 (black filled square; the data were obtained from \citealt[][]{Bruni18} and \citealt[][]{Sokolovsky17}). Regarding the BZCAT sample, this source is located in the region in-between the radio-loud and radio-weak subsamples. Also, it is located in the region where HBL and LBL sources overlap. Our RWBL-candidate locates toward the region of radio-weaker and HBL sources.

These two plots show the risk of using $R$ to set the radio loudness of blazars, since almost the entire sample would be classified as radio-loud, even the already confirmed RWBL J154419-064913. This result shows that $R$ and $L_{1.4GHz}$ are not self-consistent in order to distinguish radio-loud from radio-weak blazars. In fact, $R$ is used to distinguish radio-loud quasars from radio quiet quasars. In quasars, radio and optical emission arise from different components (the jet and the accretion disk, respectively), and are attributed to different physical processes, which is not the case for blazars. It is, however, straightforward to compare $L_{1.4GHz}$ for sources of a given class. It is then understandable that both criteria measure different situations. We calculated the mean values of $\log(R)$ for each group in the D01 sample, namely: $1.85$ for HBLs, $2.97$ for LBLs, and $3.62$ for FSRQs. 

It is worth noting that the term "radio weak" should not suggest the total lack of radio emission, but rather the intrinsically lower radio emission with respect to the general blazar population. As can be seen in Fig. 4, \RWBL\ shows intrinsically less radio power than other sources of similar characteristics (i.e., $\gamma$-ray emitting BL Lac sources), which is on itself challenging to explain using only standard spectral models (see Sect. 4).

It is straightforward to note that according to the $R$ parameter and the blazars classification, \RWBL\ remains as an HBL independently of whether the $B$ or $g'$ flux was used. We note that the HBL subsample of D01 shows a mean redshift value of $\langle z \rangle = 0.249$.

\subsubsection{A radio vs. X-ray comparison}

If the optical flux in Eq.~\ref{eq:R} is replaced by the X-ray flux (2--10 keV), another characteristic parameter arises, $R_X$, as defined by \citet[][]{Terashima03}:

\begin{equation}\label{eq:Rx}
R_X = \frac{\nu f_{5 \, \mathrm{GHz}}}{f_{(2-10 \, \mathrm{keV} )}}.
\end{equation}

\noindent The advantage of $R_X$ over $R$ is that the extinction that may affect the optical flux measurements should not be critical in the X-ray flux. In order to calculate $R_X$ for the BZCAT samples we have used the {\it Swift} XRT data, so we have slightly changed the X-ray range to 0.3--10~keV. 

\begin{figure*}

    \includegraphics[width=0.515\textwidth]{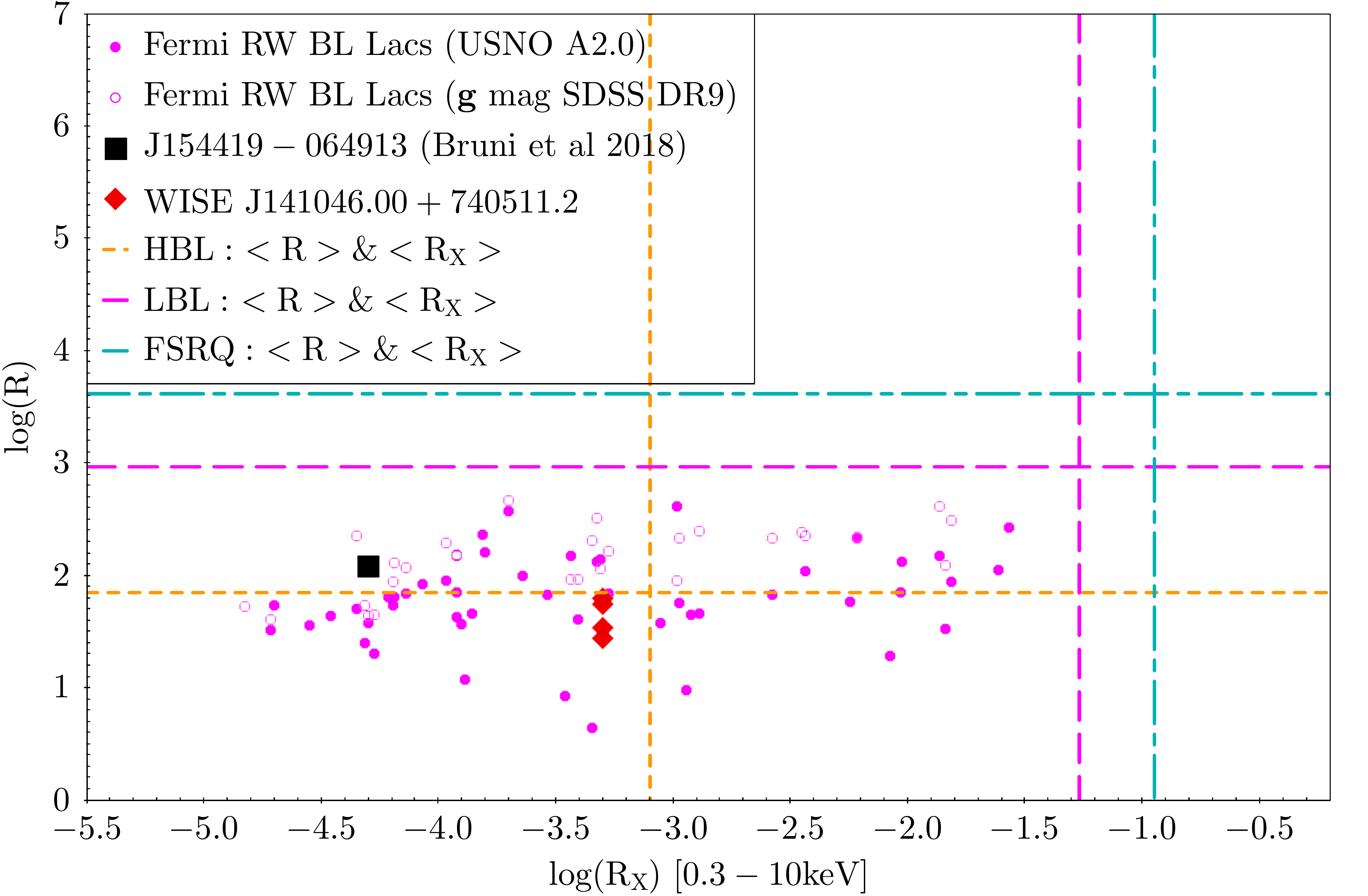}\includegraphics[width=0.515\textwidth]{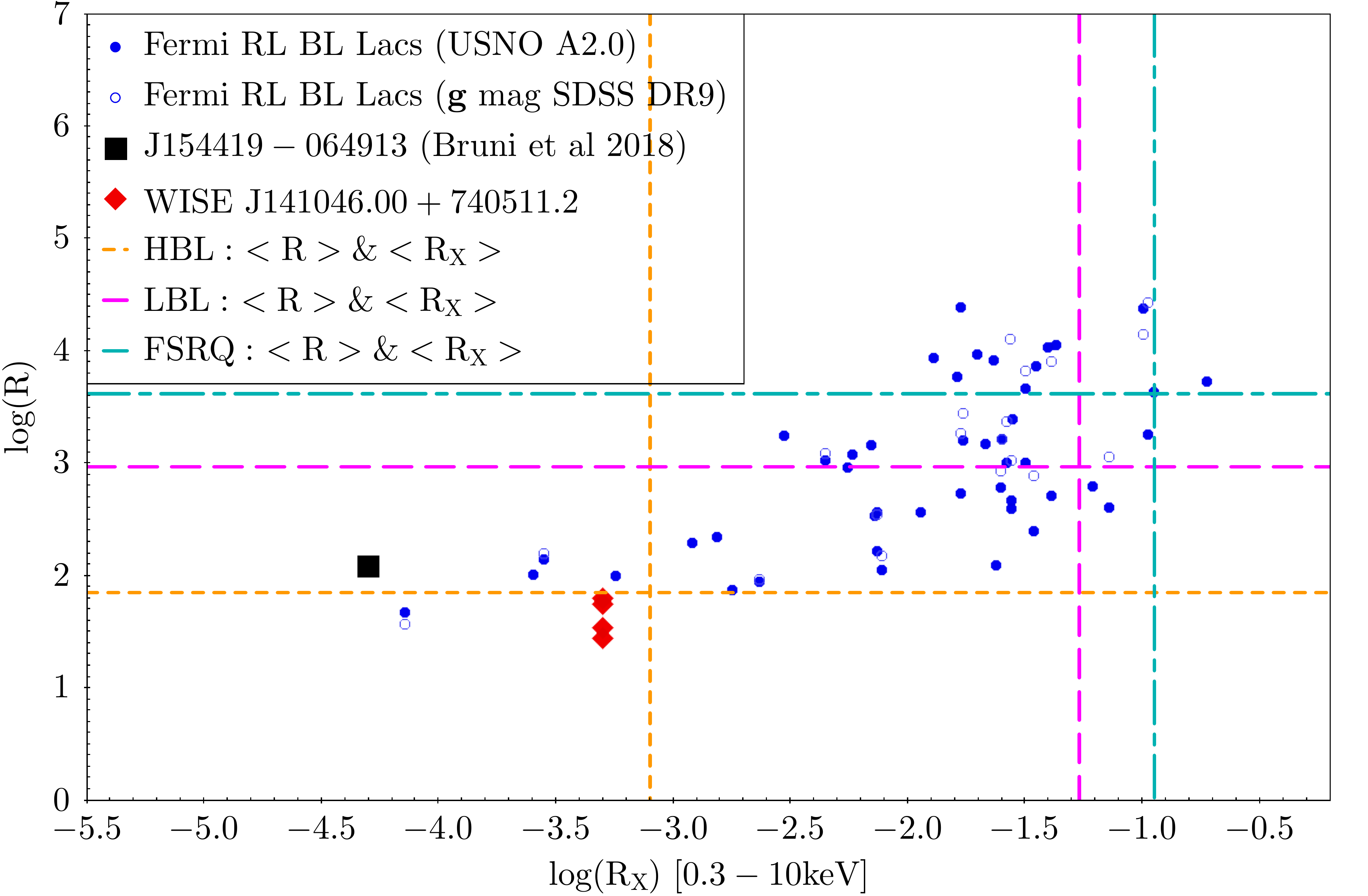}
        \centering
    \includegraphics[width=0.7\textwidth]{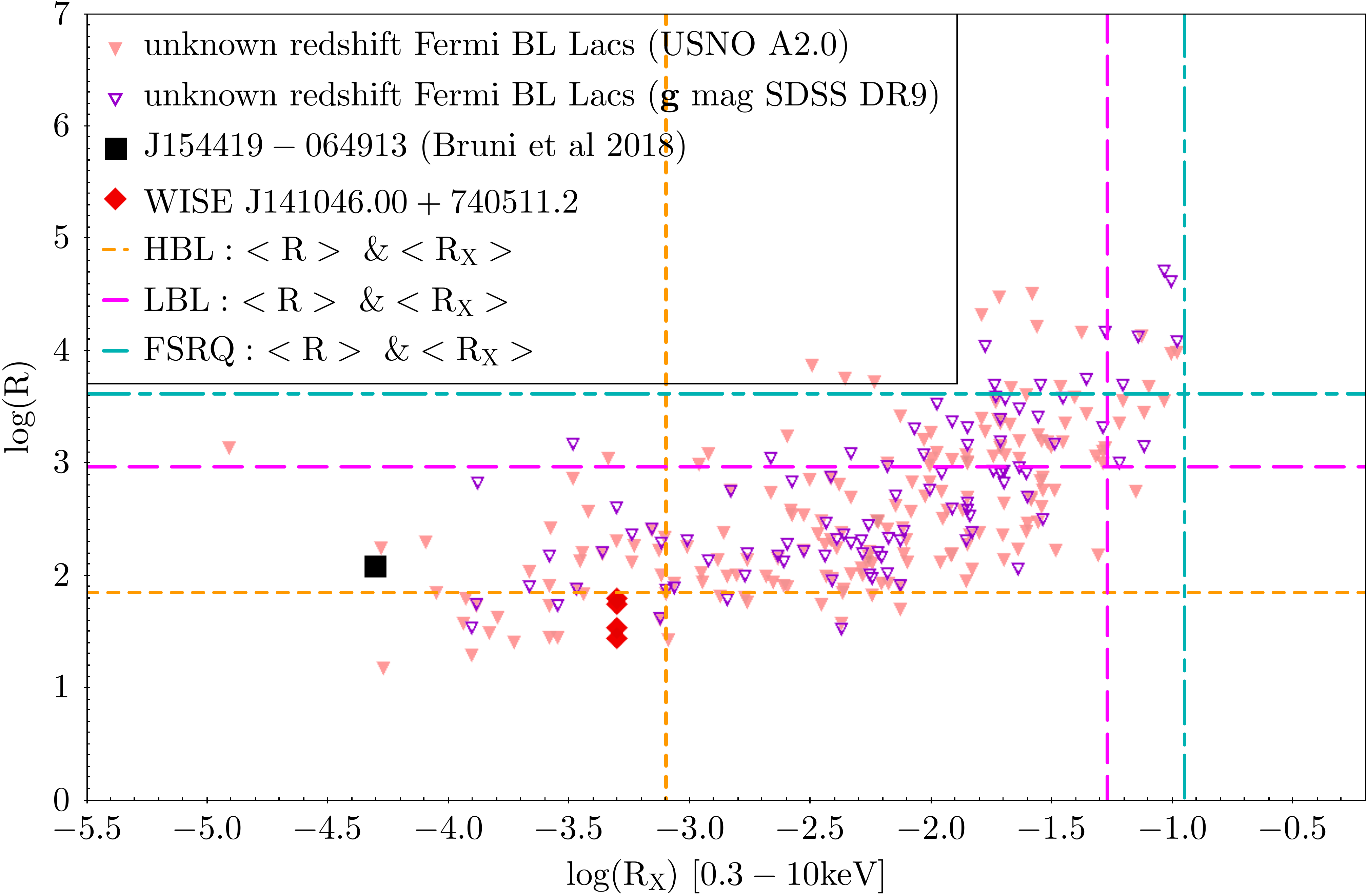}
\caption{$R$~vs.~$R_X$ diagrams. \RWBL~ is shown as the (red) filled diamonds to point to its four possible $R$ values (see Sect. 3.2.2.). The three populations of blazars are highlighted through their mean values of $R$~and $R_X$: HBLs in orange short-dashed lines, LBLs in pink long-dashed lines, and FSRQs in cyan dot-dashed lines. {\it Upper panel}: BZCAT subsamples, as in Figs.~\ref{fig:Lumin}-\ref{fig:R_z}. {\it Lower panel}: BL~Lac sources from BZCAT whose redshift, as that of our source, remain unknown.}
\label{fig:Rx}
    \end{figure*}

Figure~\ref{fig:Rx} shows the diagrams of $R$~vs.~$R_X$. The upper panels show the BZCAT radio-loud (right) and radio-weak (left) samples, separately. In the lower diagram we have used the set of BZCAT BL~Lac objects whose redshifts remain unknown, as in the case of \RWBL. As in Fig.~\ref{fig:R_z}, open symbols are used to highlight that $R$ was calculated by using the optical {\it g'} flux instead of {\it B} (filled symbols). The red filled diamonds point to our RWBL-candidate through its four possible $R$ values. The already known RWBL source is shown with (black) filled squares across the plots. The lines crossing the plots represent the mean values of $R$ and $R_X$ for HBLs (orange), LBLs (pink), and FSRQs (cyan).

The two upper panels show that the radio-weak and radio-loud samples tend to be located in rather different zones of the diagram; there is a clear overlapping region, also seen in other diagrams. The RWBL and our candidate are located toward the region of radio-weak sources. However, more importantly, these diagrams highlight that different populations of blazars occupy different zones in the $R_X$~vs.~$R$ diagram, especially regarding HBLs with respect to both LBLs and FSRQs. In this sense, both J154419$-$064913 and \RWBL\  should be assumed to be HBLs. The lower diagram of Fig.~\ref{fig:Rx} shows that BL~Lac sources with unknown redshift could belong to any of the three subpopulations and that many of these sources behave in the same way as \RWBL\ and J154419$-$064913 in the $R$~vs.~$R_X$ space. This plot deserves further analysis to evaluate whether or not it can become a useful tool to distinguish not only between HBL and LBL populations, but also between radio-weak and radio-loud sources. 

Indeed, we note that the $R_X$ parameter is closely related to the $\Phi_{XR}$ parameter used in \citet{Maselli10a}:

\begin{equation}\label{eq:Phi}
 \Phi_{XR} = 10^{-3} \frac{F_{X}}{S_{\rm 1.4~GHz} \Delta \nu}.
\end{equation}

\begin{figure*}
\centering
    \includegraphics[width=0.7\textwidth]{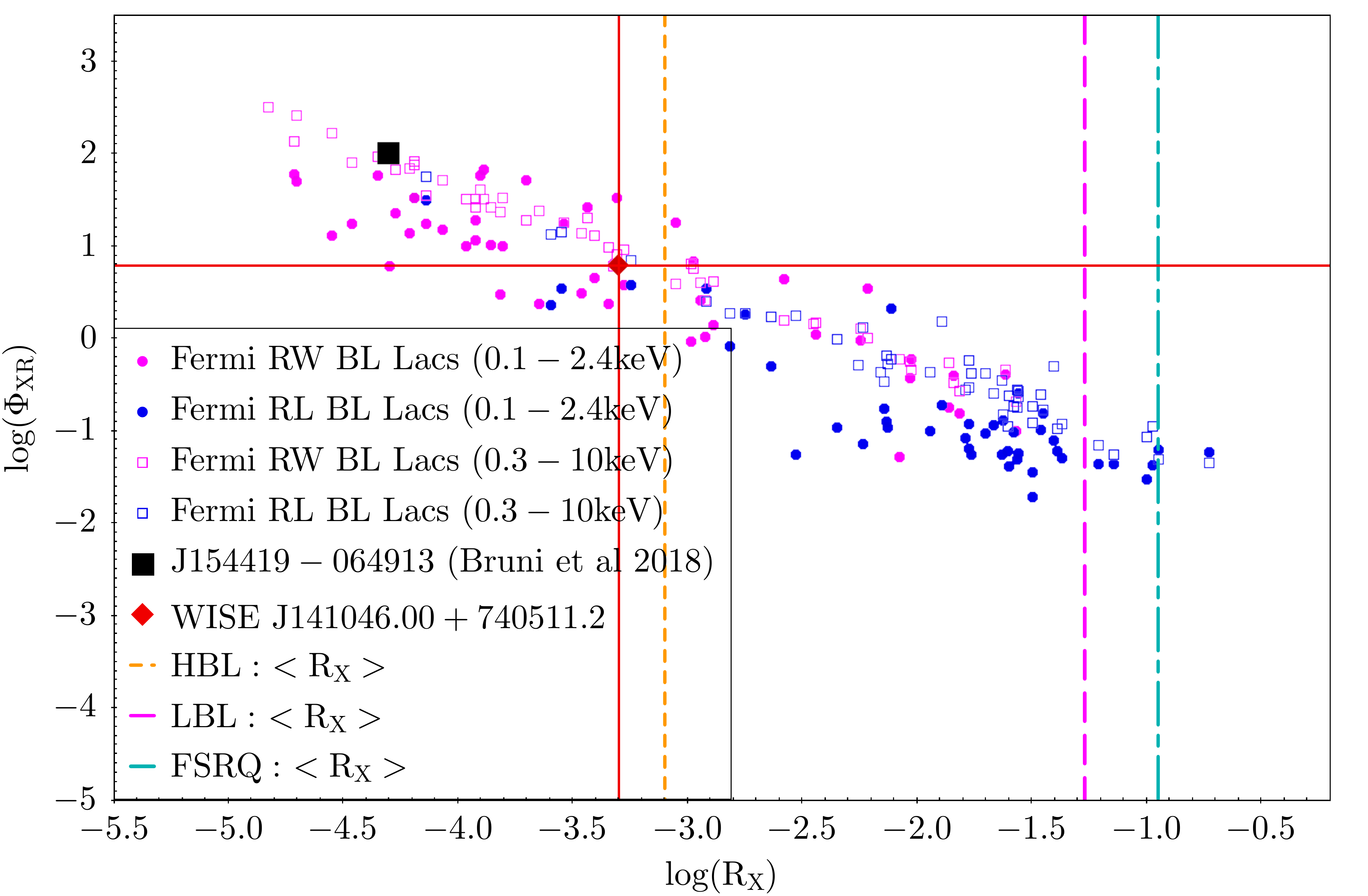}
\caption{Relationship between $\Phi_{XR}$ and $R_X$. Both the RWBL J154419$-$064913 and our RWBL-candidate are located within the region of HBL sources, according to $\Phi_{XR}$, marked with black and red lines, respectively.}
\label{fig:Phi}
    \end{figure*}

\noindent However, the parameter $\Phi_{XR}$ was built to distinguish between LBLs and HBLs. So, among all the available parameters in the literature (and also those used in this work), this is the first one developed by using blazars. We have plotted both $R_X$ and $\Phi_{XR}$ in Fig.~\ref{fig:Phi}. As in previous diagrams, we  used the BZCAT subsamples, but different symbols (circles and squares) are adopted in order to highlight that different X-ray bands were taken into account in the calculus of $\Phi_{XR}$. We note that the flux of \RWBL~ was obtained in the 0.5--10~keV band. According to \citet{Maselli10a}, $\log(\Phi_{XR})=0$ sets the threshold above which HBL sources are expected to be found. Both \RWBL~ and J154419$-$064913 locate in that region.

We conclude that the criterion adopted by \citet[][]{gregg-1996} is a straightforward definition of radio power for blazars, although it relies on knowing the distance to the source, which for BL Lac objects may prove challenging. On the other hand, the $R_X$ parameter, which is in turn closely related to the HBL/LBL spectral classification, may prove to be a viable way to predict the intrinsic radio power of a source. The HBL/LBL classification for sources with known distances is related to their radio power. The use of mixed X-ray and radio criteria has been proposed recently as a means of finding new blazars \citep[][]{Marchesini20}. Both the aforementioned methods unambiguously classify \RWBL\ as a radio weak source. The $R$ parameter, instead, is not representative of the gamma-ray parent population of BL Lac sources, which is due to the fact that it was originally defined for quasars.

\section{Modeling the broadband spectral energy distribution of \RWBL}

In order to investigate the emission properties of this source, we applied a leptonic one-zone model, based on \citet{Tavecchio98} and \citet{Ghisellini09}. Basically, we considered that the emission is produced in a spherical blob of radius, $R_{\rm b}$, that moves relativistically with bulk Lorentz factor $\Gamma$. The blob is located at a dissipation distance from the black hole, $z_{\rm diss}$,  where the magnetic field is considered to be uniform and, in the comoving frame, given by:

\begin{equation}
B'(z_{\rm diss}) = \frac{1}{\Gamma} \sqrt{ \frac{ \xi 8 \pi L_{\rm j}}{c R_{\rm b} } },\end{equation}

\noindent where $\xi$ is a free parameter of the model, accounting for the ratio between the magnetic to jet kinetic energy density, and $L_{\rm j}$ is the jet luminosity. We adopted $R_{\rm b} = 0.1z_{\rm diss}$, which is consistent with a conical jet with an opening angle of $\sim 0.1$ (equivalent to $\sim 6^{\circ}$). 

At the location of the blob, particles can be accelerated up to relativistic energies. Hence, we consider that a fraction $\eta$ of the jet kinetic energy density goes into non-thermal electrons, with an injection function represented by a power-law with index $\alpha$. The particle distribution in the blob is obtained by solving a transport equation, taking into account the radiative cooling. In all cases, we considered $\eta = 0.1$, and the minimum Lorentz factor of the electrons, $\gamma^e_{\rm }$, is a free parameter of the model.

As discussed in the previous sections, all data seem to indicate that this is an HBL, hence, we do not include any external photon source for IC scattering \citep{Ghisellini09}. The optical spectra do not show features that would be evident in the presence of an external component (i.e., clouds). Thus, we chose to consider a pure SSC model. However, it must be noted that such component could still be present if its emission is swamped by the emission arising from the relativistic jet aligned with the line of sight.

Given the amount of free parameters, the model is clearly degenerated. In order to constrain the degeneracy, we fixed some of the parameters to standard values for HBLs: we considered a black hole mass of $M_{\rm BH} = 10^8 M_{\odot}$, and a bulk Lorentz factor of $\Gamma =10$; in addition, for an inclination angle of the jet with respect to the line of sight of  $\theta_{j} = 1/ \Gamma$, we obtained a Doppler factor $\delta \sim \Gamma$. Then, we varied the free parameters within the following ranges: $L_{j} = 10^{43-45}$ erg s$^{-1}$, $z_{\rm diss} = 500-2500 r_{\rm g}$ (with $r_{\rm g}=GM/c^2$ being the gravitational radius), $\xi=10^{-6}-1$, $\alpha = 1.5-3$, and $\gamma^e_{\rm }= 1 - 10^4$.

\begin{figure*}
    \centering
    \includegraphics[width=0.80\textwidth]{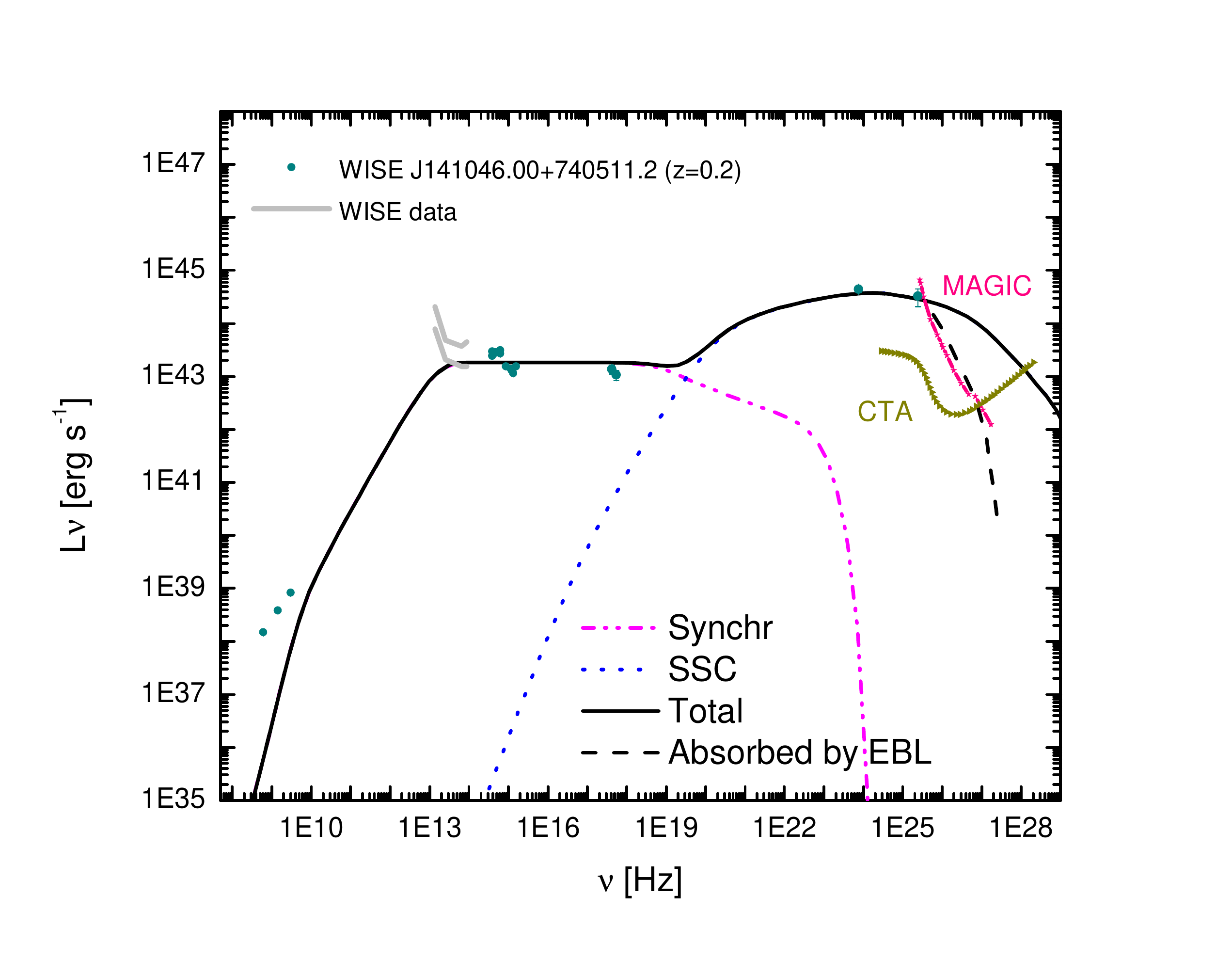}
    \includegraphics[width=0.80\textwidth]{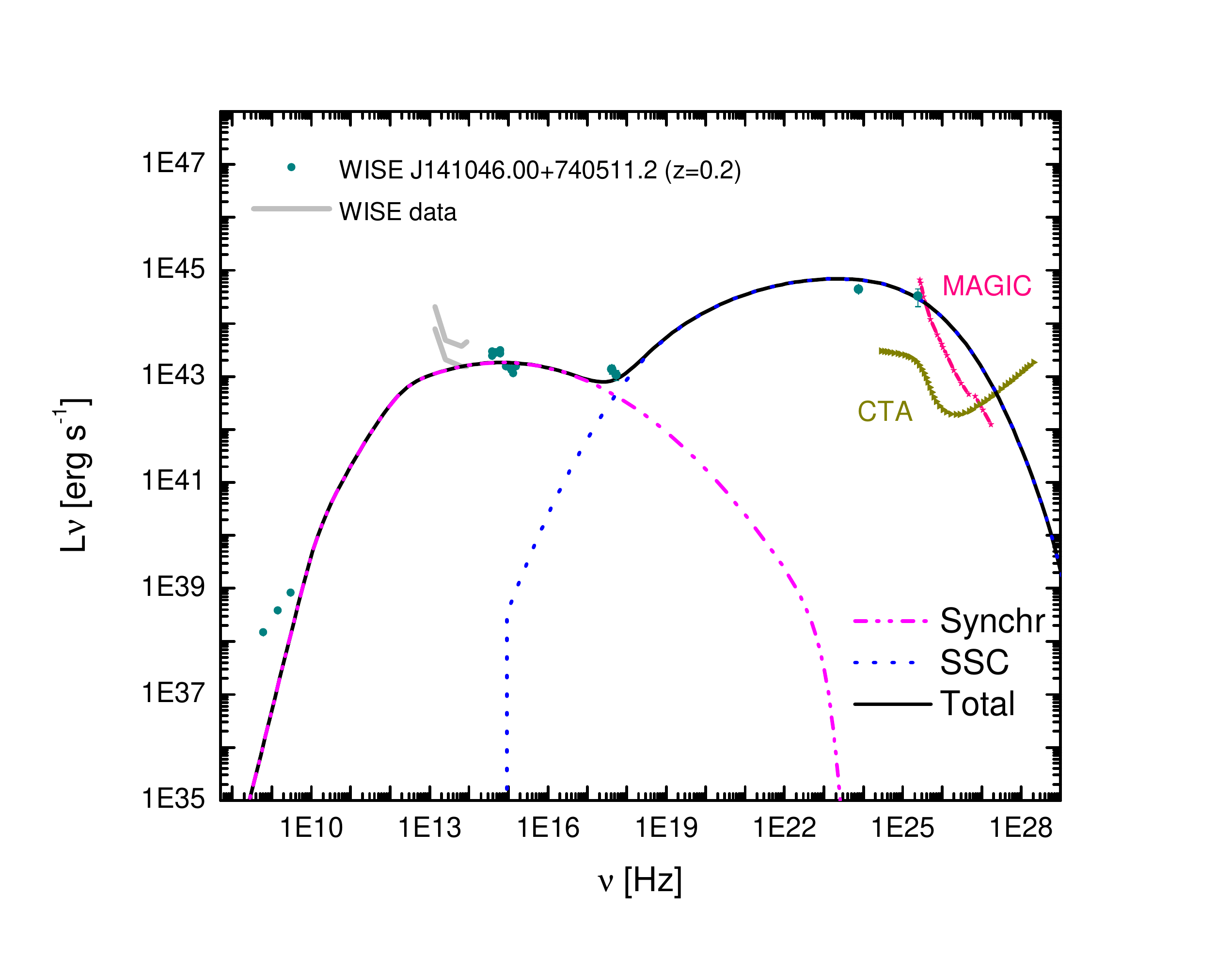}
\caption{Broadband SED of \RWBL. The solid circles represent the simultaneous multiwavelength data of the source. Data from WISE, which are not contemporaneous, are shown in gray curves (maximum and minimum fluxes). The black solid line is the SED obtained with the leptonic one-zone model, using the best-fit parameter for the power-law distribution (top panel) and the log-parabolic one (bottom panel); dashed black line shows the SED absorbed by the EBL \citep{dominguez2011}. The sensitivity curves of MAGIC \citep{MAGIC2016} and CTA are also included.}
\label{fig:fit}
    \end{figure*}

Following \citet{Massaro2004,Massaro17}, we also considered a log-parabolic
model for the particle distribution, given by:

\begin{equation}
N(\gamma) \propto (\gamma/\gamma_0)^{-(s+r\log(\gamma/\gamma_0))},
\end{equation}

\noindent where $\gamma_0$ is a reference energy, $s$ is the spectral index at the reference energy, and $r$ the curvature of the parabola, that is, the spectral curvature.
In order to fit the data with this model, we use the online tool for generating AGN SEDs\footnote{\url{https://www.isdc.unige.ch/sedtool/PROD/SED.html}} \citep{massaro06,Tramacere07,tramacere2009,tramacere2011}. We found a good agreement between both the power-law and the log-parabolic distributions.
Fig. \ref{fig:fit} shows the broadband SED obtained with the best-fit parameters for the two models for particle distributions, for a redshift of $z=0.2$. Similar plots were obtained for $z=1$.

We included the WISE data in the plot for completeness, since infrared fluxes are relevant to the proper classification of a blazar \citep[][]{DAbrusco12,Massaro16}. However, we note these fluxes should be taken with caution, since they are not contemporaneous with those at other frequencies. In addition, the source showed a highly variable behavior in the WISE bands W1 and W2, with changes up to an order of magnitude during the observation. Taking these points into account, we see that the emission in our model underestimates the far-IR data; this excess might be explained by emission in extended regions of the jet \citep{Valverde2020}.

Table \ref{tab:fit} summarizes the parameter values for the power-law model. We obtained: $B = 0.03$ and $0.7$ G, for $z=0.2$ and $z=1$, respectively, and $\gamma^e_{\rm }= 10^3$ for both cases.
Using these same conditions of the emitting regions, we went on to vary the parameter of the log-parabolic distribution, obtaining the following parameters: $s=0.9$, $r=0.32$, and $\gamma_0=9$.

 \begin{table}[ht]
  \caption[]{Best-fit parameter values.}
     \label{tab:fit}
     \centering
         \begin{tabular}{lcc}
            \hline
             \hline
             Parameter & $z=0.2$ & $z=1$ \\
            \hline
            $z_{\rm diss}/r_{\rm g}$ & 1000 & 500 \\
            $L_{j}$ [erg s$^{-1}$] & $10^{45}$ & $10^{45}$ \\
            $B$  & 0.03 G & 0.7 G \\
            $\alpha$ & 3 & 2.5 \\
            $\gamma^e_{\rm }$& $10^3$ & $10^3$ \\
            \hline
         \end{tabular}
     
   \end{table}
   
For both values of $z$ and for both models of the particle distribution used to reproduce the data, the synchrotron peak is beyond $10^{17}$ Hz, probably suggesting that \RWBL\ is an EHBL \citep{Ghisellini98}. Our results are in accordance with those found in the literature studying the behavior of EHBLs, in particular, magnetic fields well below equipartition \citep{Bonnoli15,acciari2020} and display high values of minimum Lorentz factor of electrons \citep{Tavecchio09,Bonnoli15}. In addition, the weak radio flux detected for this source is in accordance with this classification, as discussed in \citet{Bonnoli15}. Nevertheless, the classification of this object as an EHBL is still model-dependent and requires further analysis. 

The size of the emitting region and the Doppler factor are in agreement with a variability timescale of $\sim 1$ d (see \cite{Massaro17} for a mid-IR variability analysis). Given the compactness of the region ($R_{\rm b} ~ 10^{  15}$ cm), synchrotron radiation is self-absorbed at radio frequencies \citep{Tavecchio98}. Since the source is not resolved by VLA observations, there is a maximum size of the radio emitting region compatible with these results. Considering redshifts of $z=0.2/1$, this region should be of $\lesssim$~kpc scales. In a forthcoming work, we will study radio emission produced in more extended regions of the jet \citep[e.g.,][]{Ghisellini05}, taking into account these constraints with the aim of  explaining the radio weakness of this HBL.

\section{Summary and conclusions}\label{conc}

We  conducted a multiwavelength observational campaign  of the source \RWBL\  to study its nature. Originally detected by Fermi, it was later revealed to show several BL Lac characteristics in the $\gamma$-ray, X-ray, UV, optical, and infrared bands, but its radio emission was not on par with what is expected for a radio-loud source, as a canonical BL Lac would demonstrate.

We state that \RWBL\ is indeed a $\gamma$-ray emitting BL Lac source, given its typical BL Lac optical spectrum as observed by high-sensitivity telescopes and its observed high degree of optical polarization. To estimate an intrinsic polarization degree, the linear polarization of the synchrotron flux, produced by an electron power-law energy spectrum $ N(E) \propto E^{-\Gamma}$ in an optically thin source with a uniform magnetic field $B_0$, is given by \citep[e.g.,][]{Pacholczyk1970}:

\begin{equation}
P_0(\Gamma) = \frac{3\Gamma+ 3}{3\Gamma+ 7}.
\end{equation}

\noindent For an index of $\Gamma=2.5-3$, namely, the values obtained for the electrons, we derive an intrinsic polarization of $P_0\sim 72-75$ \%. Even in the worst-case scenario conditions (such as a magnetic field structure as in the launching region of the jet, \citet{vieyro2016}), the intrinsic polarization would remain above $P_0\sim 75$ \%. This is the strongest evidence that this source is of the BL Lac kind.

We also claim that \RWBL\ most probably lies at a moderate redshift ($z\lesssim0.2$), since it would agree with both the observed data (no spectral features detectable in the 400-1000 nm range, and the fraction of radio and X-ray emission to optical flux), and the theoretical emission models.
\RWBL\ has been detected in radio frequencies, with a flux $< 2.5$ mJy, making it one order of magnitude less bright in radio than the rest of the Fermi BL Lacs. We classified the source as an HBL both by the fraction of its X-ray to radio flux and by the position of its synchrotron peak in SED models.

Finally, we applied a one-zone model to estimate the shape of its SED. The best-fit parameter set results in synchrotron peaks above $10^{17}$ Hz, making \RWBL\ an EHBL candidate. Some of the parameter values are also in accordance with those found for modeling EHBL \citep{Bondi01,acciari2020}. Further analysis is needed to model the radio weakness of this source.

\begin{acknowledgements}
      The team of coauthors would like to thank the anonymous referee for their constructive and very positive feedback. F.L.V. acknowledges support from the Argentine agency CONICET (PIP 2021-0554). V.R., I.A. and S.A.C. acknowledge the support from Universidad Nacional de La Plata through grant 11/G153. P.B. and J.S. acknowledge support from ANPCyT PICT 2017-0773. E.J.M. would like to thank Dr. R. I. P\'aez for the helpful feedback. E.J.M. would like to acknowledge, on behalf of all the authors, all the observing facilities and instruments that are mentioned in the following, as well as the staff involved in data acquisition. This work is based on observations obtained at the international Gemini Observatory, a program of NSF’s NOIRLab, which is managed by the Association of Universities for Research in Astronomy (AURA) under a cooperative agreement with the National Science Foundation on behalf of the Gemini Observatory partnership: the National Science Foundation (United States), National Research Council (Canada), Agencia Nacional de Investigaci\'{o}n y Desarrollo (Chile), Ministerio de Ciencia, Tecnolog\'{i}a e Innovaci\'{o}n (Argentina), Minist\'{e}rio da Ci\^{e}ncia, Tecnologia, Inova\c{c}\~{o}es e Comunica\c{c}\~{o}es (Brazil), and Korea Astronomy and Space Science Institute (Republic of Korea). In this research we utilised data acquired by the Gran Telescopio Canarias. GTC is a Spanish initiative led by the Instituto de Astrof\'isica de Canarias (IAC). The project is actively supported by the Spanish Government and the Local Government from the Canary Islands through the European Funds for Regional Development (FEDER) provided by the European Union. The project also includes the participation of Mexico (Instituto de Astronom\'ia de la Universidad Nacional Aut\'onoma de M\'exico (IA-UNAM) and Instituto Nacional de Astrof\'isica, \'Optica y Electr\'onica (INAOE), and the US University of Florida.
      This work includes data acquired at the Giant Metrewave Radio Telescope, which is run by the National Centre for Radio Astrophysics of the Tata Institute of Fundamental Research in Pune, India. This article uses data taken by operating the Karl G. Jansky Very Large Array, a part of the National Radio Astronomy Observatory, a facility of the National Science Foundation operated under cooperative agreement by Associated Universities, Inc. This work made use of data supplied by the UK Swift Science Data Centre at the University of Leicester. We acknowledge the Liverpool Telescope, which is operated on the island of La Palma by Liverpool John Moores University in the Spanish Observatorio del Roque de los Muchachos of the Instituto de Astrof\'isica de Canarias with financial support from the UK Science and Technology Facilities Council, for the provided data. This article includes data acquired by the William Herschel Telescope, which is operated on the island of La Palma by the Isaac Newton Group of Telescopes in the Spanish Observatorio del Roque de los Muchachos of the Instituto de Astrof\'isica de Canarias. This work was partially supported by CONACyT (Consejo Nacional de Ciencia y Tecnolog\'ia) research grant 280789 (Mexico). N.C.S. acknowledge support by the Science and Technology Facilities Council (STFC), and from STFC grant ST/M001326/1. J.A.C. is a Mar\'ia Zambrano researcher fellow funded by the European Union -NextGenerationEU- (UJAR02MZ). This work received financial support from PICT-2017-2865 (ANPCyT) and PIP 0113 (CONICET). J.A.C. was also supported by grant PID2019-105510GB-C32/AEI/10.13039/501100011033 from the Agencia Estatal de Investigaci\'on of the Spanish Ministerio de Ciencia, Innovaci\'on y Universidades, and by Consejer\'ia de Econom\'{\i}a, Innovaci\'on, Ciencia y Empleo of Junta de Andaluc\'{\i}a as research group FQM-322, as well as FEDER funds. The Pan-STARRS1 Surveys (PS1) and the PS1 public science archive have been made possible through contributions by the Institute for Astronomy, the University of Hawaii, the Pan-STARRS Project Office, the Max-Planck Society and its participating institutes, the Max Planck Institute for Astronomy, Heidelberg and the Max Planck Institute for Extraterrestrial Physics, Garching, The Johns Hopkins University, Durham University, the University of Edinburgh, the Queen's University Belfast, the Harvard-Smithsonian Center for Astrophysics, the Las Cumbres Observatory Global Telescope Network Incorporated, the National Central University of Taiwan, the Space Telescope Science Institute, the National Aeronautics and Space Administration under Grant No. NNX08AR22G issued through the Planetary Science Division of the NASA Science Mission Directorate, the National Science Foundation Grant No. AST-1238877, the University of Maryland, Eotvos Lorand University (ELTE), the Los Alamos National Laboratory, and the Gordon and Betty Moore Foundation.

\end{acknowledgements}

%
%

\newpage
\bibliography{Biblio} 
\bibliographystyle{aa}
\newpage

\end{document}